\documentclass[12pt]{article}
\usepackage{amsmath}
\usepackage{amsthm}
\usepackage{amscd}
\usepackage{epsfig}
\usepackage{textcomp}
\usepackage{fullpage}
\usepackage{natbib}
\usepackage{setspace}
\usepackage{amsfonts}
\usepackage{color}

\newcommand{\bm}[1]{\mbox{\boldmath $#1$}}
\newcommand{\mb}[1]{\mathbf{#1}}

\newcommand{\NA}[0]{\mbox{\tt NA}}
\newcommand{\ith}[1]{$#1^{\mbox{\tiny th}}$}
\DeclareMathOperator*{\argmin}{argmin}

\begin{document}

\title{
On estimating covariances between many assets
  with histories of highly variable length}
\author{
  Robert B. Gramacy\\
  Statistical Laboratory\\
  University of Cambridge\\
  bobby@statslab.cam.ac.uk \and
  Joo Hee Lee \\
  Fidelity Investments \\
  London\\
  joohee.lee@uk.fid-intl.com \and
  Ricardo Silva\\
  Department of Statistical Science\\
  University College London\\
  ricardo@stats.ucl.ac.uk
}

\maketitle

\doublespacing

\begin{abstract}
  Quantitative portfolio allocation requires the accurate and
  tractable estimation of covariances between a large number of
  assets, whose histories can greatly vary in length.  Such data are
  said to follow a monotone missingness pattern, under which the
  likelihood has a convenient factorization.  Upon further assuming
  that asset returns are multivariate normally distributed, with
  histories at least as long as the total asset count, maximum
  likelihood (ML) estimates are easily obtained by performing repeated
  ordinary least squares (OLS) regressions, one for each asset. Things
  get more interesting when there are more assets than historical
  returns.  OLS becomes unstable due to rank--deficient design
  matrices, which is called a ``big $p$ small $n$'' problem.  We
  explore remedies that involve making a change of basis, as in
  principal components or partial least squares regression, or by
  applying shrinkage methods like ridge regression or the lasso.  This
  enables the estimation of covariances between large sets of assets
  with histories of essentially arbitrary length, and offers
  improvements in accuracy and interpretation.  We further extend the
  method by showing how external factors can be incorporated.  This
  allows for the adaptive use of factors without the restrictive
  assumptions common in factor models.  Our methods are demonstrated
  on randomly generated data, and then benchmarked by the performance
  of balanced portfolios using real historical financial returns. An
  accompanying {\sf R} package called {\tt monomvn}, containing code
  implementing the estimators described herein, has been made freely
  available on CRAN.

\end{abstract}

\section{Introduction}
\label{sec:intro}

Missingness in data, and hence the quest if one should eliminate a
part of the data or try and estimate characteristics of it, is common
in statistical analysis. The missing observation problem varies in
style, depending on the type of data. One example is random
missingness, which may stem from erroneous data
\citep{dempster:laird:rubin:1977}.  In financial returns data
analysis, however, one problem stands out, which we will refer to as
monotone missingness. This happens when the assets of interest have
different lengths of historical financial data, e.g., stock prices and
returns.  There are several possible ways of dealing with this type of
incomplete dataset.  One way is by utilizing the portion of data
available across all of the assets.  Another approach involves
estimating the missing portion, called {\em imputation}
\citep[e.g.,][]{little:rubin:2002}.  A third approach is the focus of
this paper.

Aside from some glitches in data, which will typically give rise to
unrealistic spikes or random missingness in data, the monotone style
of missingness that permeates financial historical returns data can be
grouped into two patterns.  The first is where the histories of assets
differ due to the fact that they have started being publicly traded at
different times. The second is where assets close for various reasons,
including corporate actions such as M\&A (Merger and Acquisition)
activities, or liquidation due to bankruptcy. Both are critical
problems to address when conducting a multivariate analysis.  In this
paper, we shall focus mainly on the former. This is sensible for the
application to portfolio balancing that we have in mind, since one is
naturally restricted to purchasing shares of companies which have
survived up to current point in time.  The latter type of missingness,
in absence of the former, can be handled similarly, but it is not
immediately clear how this would be useful for portfolio balancing.
Handling both types of monotone missingness jointly, and other types
of approximately monotone missingness, requires the method of data
augmentation \citep{schafer:1997,little:rubin:2002}. This could
potentially be useful for a descriptive analysis, but is beyond the
scope of this paper.

Data with arbitrary missingness patterns typically require specialized
iterative (even stochastic) estimation algorithms that can be slow and
cumbersome to implement.  However, data which follow a monotone
missingness pattern lead to a likelihood which has a convenient
factorization.  If we further assume that asset returns are
multivariate normally distributed (MVN), with histories at least as
long as the total asset count, then maximum likelihood (ML) estimators
are easily obtained by performing repeated ordinary least squares
(OLS) regressions, one for each asset.  In the finance literature,
this approach is usually attributed to \cite{stambaugh:1997}, but it
was first described by \cite{andersen:1957} and has since been
discussed in many texts (see Section \ref{sec:monotone}).  The method
fails when there are more assets than historical returns.  In this
case the OLS regressions become unstable due to rank--deficient design
matrices.  This is sometimes called the ``big $p$ small $n$'' problem.
It has recently received much attention in the statistics community,
with ready applications in bioinformatics and genomics, for example.
In the context of estimation for data with a monotone missingness
pattern, it can severely limit applicability to cases with a small to
modest level of missingness.

In financial applications, where there may be more assets than there
are historical price observations for (some of) the assets, this
essentially means that the method cannot be applied on the full set of
assets of interest.  This paper explores remedies to this problem.  We
aim to develop a method that can be applied in settings where some
assets have histories which are shorter than the total number of
assets, and even when there are more assets than observations.  In
short, our solution involves replacing OLS with ``parsimonious
regressions'' that either make a change of basis, as in principal
components or partial least squares regression, or apply shrinkage,
like ridge regression or the lasso.  This enables the estimation of
covariances between large sets of assets with histories of essentially
arbitrary (and uneven) length.  Even in situations where OLS would
have been sufficient, we find that the more parsimonious approach can
offer improvements in accuracy and interpretation.

The parsimonious approach also motivates novel ways of exploiting {\it
  factor} information, e.g., the value--weighted market index, size,
and book--to--market factors \citep{famafrench:1993}.  Traditionally,
factor models require the restrictive assumption that assets are
independent given the factors.  This underlying assumption can be
thought of as a specific type of parsimony.  We show how one can use
the data to decide which independence constraints are reasonable, by
incorporating the factors into our proposed framework, and furthermore
how this may be accomplished even under condition of monotone
missingness in the historical returns {\em and} factors.

The remainder of the paper is organized as follows.  Section
\ref{sec:monotone} defines the monotone pattern for missing data,
derives the corresponding factorized likelihood, and gives an
algorithm of repeated regressions to analytically find a ML estimator
for the case where the sampling distribution is assumed to be MVN.
Section \ref{sec:bpsn} outlines methods for dealing with the ``big $p$
small $n$'' problem in the context of regression with transformed
inputs and shrinkage estimators.  We highlight the benefits of
increased applicability, accuracy, and interpretability obtained with
these methods.  Section \ref{sec:monomvn} gives the details of an
algorithm---for MVN data under a monotone missingness pattern---that
combines the method in Section \ref{sec:monotone} with the
parsimonious regressions in Section \ref{sec:bpsn}. We explain how the
method can easily integrate factor information, generating a model
that essentially mixes factor models with estimators that account for
the direct dependency between returns.  We then briefly describe an
implementation which has been made freely available as an {\sf R}
package called {\tt monomvn}.  Section \ref{sec:results} shows the
method in action on synthetic data and real financial data with large
numbers of assets having histories of highly varying length.  Our
results are benchmarked against several standard comparators in the
context of covariance estimation and portfolio balancing, and are
accompanied by comments on interpretation, efficiency, and on the
(benign) consequences of using a method that leverages an MVN
assumption when that assumption not believed to hold.
Finally, we conclude with a discussion in Section \ref{sec:discuss}
that focuses on some of the limitations inherent in taking a maximum
likelihood approach.

\section{Multivariate normal monotone missing data}
\label{sec:monotone}

Let $\mb{Y}$ be a $n \times m$ matrix of random observations $Y_{i,j}$
which may not be completely observed.  Denote $y_{i,j} = \NA$ if the
\ith{i} sample of the \ith{j} covariate is missing.  In other words,
if the columns of a sampled $\mb{Y}$: $y_{:,1},\dots, y_{:,m}$,
represent a historical return series of assets indexed by $j$ and a
return for asset $j$ is not available at time $i$, then $y_{i,j} =
\NA$.  Observed $\mb{Y}$ are said to follow a {\em monotone
  missingness pattern} [e.g., \citep[][Section 6.5.1]{schafer:1997} or
\citep[][Section 7.4]{little:rubin:2002}] if the columns can be
arranged so that $y_{i,j} \ne \NA$ whenever $y_{i,j+1} \ne \NA$.
\begin{figure}[ht!]
\centering
\input{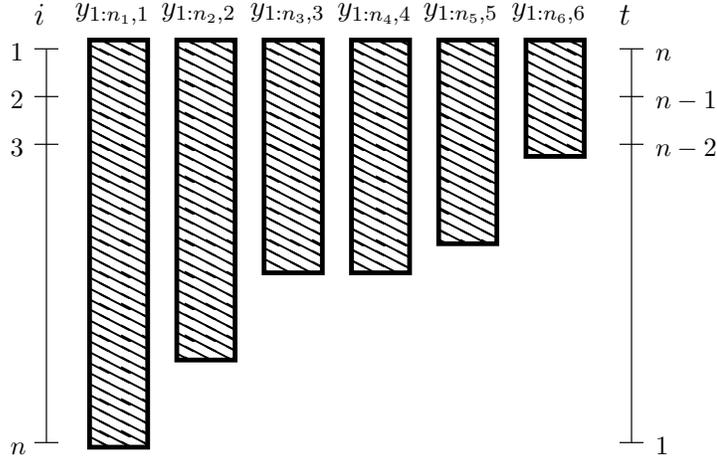}
\caption{Diagram of a monotone missingness pattern with $m=6$
  covariates, with a maximum of $n$ completely observed samples in
  $\mb{y}_1=y_{:,1}$.}
\label{f:mono}
\end{figure}
Figure \ref{f:mono} illustrates this property diagrammatically.  The
row dimension $n$, of $\mb{Y}$, is equal to the number of completely
observed samples $n_1$ of $\mb{y}_1 \equiv y_{:,1}$, the maximally
observed column.  Similarly, let $\mb{y}_j \equiv y_{1:n_j,j}$ collect
the complete data in the \ith{j} column of $\mb{Y}$, so that $n_j \geq
n_{j+1}$.

The monotone missingness patterns considered in this paper are assumed
to be {\em missing completely at random} (MCAR) in that the pattern of
missingness neither depends on the observed nor unobserved responses.
Note that there may be columns with identical missingness patterns.
In the case of asset return series with observed histories going back
different amounts of time, the MCAR assumption may be tenuous, but it
is commonly asserted anyway \citep[e.g.,][]{stambaugh:1997}.  In our
notation, the time index ($t$) for an asset's return history would run
counter to $i$, the index of the rows of $\mb{Y}$; i.e, $t=n-i+1$, as
also illustrated in Figure \ref{f:mono}.

When the missing data pattern is monotone, the likelihood $f(\mb{Y}|
\bm{\theta})$ can generally be factorized by exploiting an auxiliary
parameterization $\bm{\phi}=(\bm{\phi}_1, \dots, \bm{\phi}_m)$:
\[
f(\mb{Y}|\bm{\theta}) = f(\mb{y}_1|\bm{\phi}_1)
f(\mb{y}_2|\mb{y}_1,\bm{\phi}_2)
f(\mb{y}_3|\mb{y}_1,\mb{y}_2,\bm{\phi}_2) \cdots f(\mb{y}_m |
\mb{y}_1,\dots,\mb{y}_{m-1},\bm{\phi}_m).
\]
together with a mapping $\bm{\phi} = \Phi(\bm{\theta})$.
With the appropriate conditioning, the $y_{i,j}$ are assumed to be
independent and identically distributed (i.i.d.), so that
\begin{equation}
f(\mb{y}_j | \mb{y}_1,\dots \mb{y}_{j-1}, \bm{\phi}_j) = \prod_{i=1}^{n_j} 
f(y_{i,j}|y_{i,1}\dots, y_{i,j-1}, \bm{\phi}_j). \label{eq:iidlik}
\end{equation}
We are concerned with the case where the $(y_{i,1},\dots y_{i,m})$
follow a multivariate normal distribution (MVN) so that the likelihood
in (\ref{eq:iidlik}) also follows a MVN with constant variance and a
mean linear in $y_{i,1},\dots, y_{i,j-1}$.  The i.i.d.~and MVN
assumptions may be less than ideal for financial returns data
\citep[e.g.,][]{mills:1927}, but we note that these are common
simplifying assumptions \citep{stambaugh:1997,ckl:1999,jagma:2003}
because they lead to tractable inference and compare favorably (see
Section \ref{sec:results} for results and further discussion).
Maximum likelihood estimators (MLEs) of $\bm{\theta}_j = (\mu_j,
\bm{\Sigma}_{1:j,j})$, $j=2,\dots,m$, can then be obtained by
regression on the complete data:
\begin{align}
  \mb{y}_j &= \mb{Y}_j \bm{\beta}_j + \bm{\epsilon}_j, &
  \{\epsilon_{i,j}\}_{i=1}^{n_j} &\stackrel{\mbox{\tiny i.i.d.}}{\sim}
  N(0,\sigma_j^2) \label{eq:monoreg}
\end{align}
where $\bm{\beta}_j^\top = (\beta_{0,j}, \beta_{1,j}, \dots,
\beta_{(j-1),j})$ and $\mb{Y}_j \equiv \mb{Y}_{0:(j-1)}^{(n_j)}$ is
the $n_j \times j$ design matrix
\[
\mb{Y}_j \equiv \mb{Y}_{0:(j-1)}^{(n_j)} = \begin{pmatrix}
  1 & y_{1,1} & \cdots & y_{1,(j-1)} \\
  1 & y_{2,1} & \cdots & y_{2,(j-1)} \\
  \vdots & \vdots & \ddots & \vdots \\
  1 & y_{n_j,1} & \cdots & y_{n_j, (j-1)}
\end{pmatrix}
\]
containing an intercept column, and the first $n_j$ observations of
the first $j-1$ columns of $\mb{Y}$.  So the auxiliary parameters
used in (\ref{eq:monoreg}) are $\bm{\phi}_j = (\bm{\beta}_j,
\sigma_j^2)$.
\begin{figure}[ht!]
\centering
\input{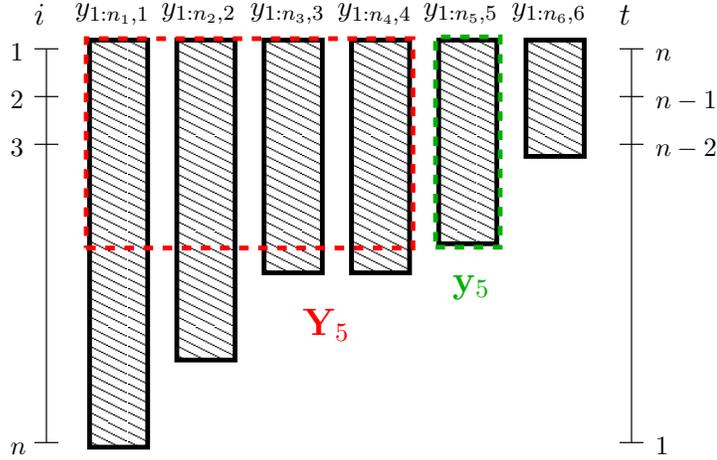}
\caption{Diagram of the design matrix $\mb{Y}_5$ (without an intercept
  term) and the response vector $\mb{y}_5$ for the fifth regression
  involved in maximizing the likelihood of MVN data under a monotone
  missingness pattern with $m=6$ covariates.}
\label{f:monoreg}
\end{figure}
Figure \ref{f:monoreg} diagrams the design matrix (without the
intercept term) and response vector involved in one such regression.
When $\mathrm{rank}(\mb{Y}_j) = j$, and particularly when $n_j > j$,
MLEs $\hat{\bm{\phi}}_j$ are obtainable via the straightforward
calculation:
\begin{align}
\hat{\bm{\beta}}_j &= (\mb{Y}_j^\top \mb{Y}_j)^{-1} \mb{Y}_j^\top \mb{y}_j &
\mbox{and} && 
\hat{\sigma}^2_j &= \frac{1}{n_j} ||\mb{y}_j - \mb{Y}_j \hat{\bm{\beta}}_j||^2
= \frac{1}{n_j} \sum_{i=1}^{n_j} (y_{i,j} 
- (\mb{y}_i^\top)_{1:n_j}\, \hat{\bm{\beta}}_j)^2.
\label{eq:regress}
\end{align}
Then,
starting with $\hat{\bm{\theta}}_1$ comprising of $\hat{\mu}_1 =
\sum_{i=1}^{n_1} y_{i,1}/{n_1}$, and $\hat{\Sigma}_{1,1} =
\sum_{i=1}^{n_1} (y_{i,1} - \hat{\mu}_1)^2/{n_1}$, each
$\hat{\bm{\theta}}_j$ can be estimated conditional on
$\hat{\bm{\theta}}_{1:(j-1)} = (\hat{\bm{\mu}}_{1:(j-1)}^\top,
\hat{\bm{\Sigma}}_{1:(j-1),1:(j-1)})$ and estimates of
$\hat{\bm{\beta}}_j$ and $\hat{\sigma}^2_j$ as \citep{stambaugh:1997}:
\begin{align}
  \hat{\mu}_j &= \hat{\beta}_{0,j} + \hat{\bm{\beta}}_{1:(j-1),j}^\top
  \hat{\bm{\mu}}_{1:(j-1)}
&\hspace{-0.075cm} \mbox{and}&&
\hat{\bm{\Sigma}}_{1:j,j}
  &= \begin{pmatrix}
    \hat{\bm{\beta}}_{1:(j-1),j}^\top \hat{\bm{\Sigma}}_{1:(j-1),1:(j-1)} \\
    \hat{\sigma}^2_j + \hat{\bm{\beta}}_{1:(j-1),j}^\top
    \hat{\bm{\Sigma}}_{1:(j-1),1:(j-1)} \hat{\bm{\beta}}_{1:(j-1),j},
\label{eq:addy}
\end{pmatrix}
\end{align}
thus implicitly describing the mapping $\Phi^{-1}$ back to
$\bm{\theta}_j$--space.  Observe that we do not use a bias--corrected
estimator for $\sigma_j^2$ in (\ref{eq:regress}), i.e., with $n_j-j$
instead of $n_j$ in the denominator, to ensure that ML estimates
$\hat{\bm{\theta}}$ are obtained \citep[][pp.~224]{schafer:1997}.
However, we have found it to be beneficial in practice to use $n_j-1$
in the denominator as is typical in obtaining unbiased estimates of
covariance matrices in the complete data case.

When several columns $\mb{y}_\ell$, say $\ell=j_1,\dots,j_2$, have
equal lengths of observed histories $n_\ell$, it is typical to use a
multivariate regression $(\mb{y}_{j_1} \; \cdots \; \mb{y}_{j_2}) =
\mb{Y}_{j_1} \bm{\beta}_{j_1:j_2} + \bm{\epsilon}_{j_1:j_2}$ to find
$\hat{\bm{\beta}}_{j_1:j_2}$ and the empirical variance--covariance
matrix $\hat{\mb{V}}_{j_1:j_2,j_1:j_2}$.  Then, several
$\hat{\bm{\theta}}_{j_1:j_2}$ can be found at once by replacing
$\hat{\bm{\beta}}_j$ with $\hat{\bm{\beta}}_{j_1:j_2}$ and
$\hat{\sigma}_j^2$ with $\hat{\mb{V}}_{j_1:j_2,j_1:j_2}$ in
(\ref{eq:addy}).  Importantly, if
$\hat{\bm{\Sigma}}_{1:(j_1-1),1:(j_1-1)}$ and
$\hat{\mb{V}}_{j_1:j_2,j_1:j_2}$ are positive definite, then
$\hat{\bm{\Sigma}}_{1:j_2,1:j_2}$ will be positive definite as well
\citep{stambaugh:1997}.

Calculating such MLEs requires having $n_j > j$ for all $j=1,\dots,m$.
That is, there cannot be an asset whose history is shorter than the
number of assets whose histories have greater length.  If such were
the case, then $\mb{Y}_j$ would not be of full rank, and $\mb{Y}_j^\top
\mb{Y}_j$ could not be inverted in Eq.~(\ref{eq:regress}).  This is
sometimes referred to in the literature as the problem of regression
with ``big $p$ [number of parameters] small $n$ [number of
observations]''. Numerical singularities may arise whenever $n_j$ is
less than, but nearly equal to, $j$---especially when $n$ and $m$ are
large.  In the following section we illustrate how these difficulties
may be overcome by methods of subset selection, coefficient shrinkage,
or the use of principal components.

\section{Parsimonious regression}
\label{sec:bpsn}

In this section, we extract and focus on the subproblem of the linear
regression in (\ref{eq:monoreg}), in terms of a design matrix of $p$
predictor variables with an intercept term ($\mb{X} \equiv \mb{Y}_j$)
observed for $n$ cases, with corresponding responses ($\mb{y} \equiv
\mb{y}_j$, where $n \equiv n_j$):
\begin{align}
  \mb{y} &= \mb{X} \bm{\beta} + \bm{\epsilon}, &
  \{\epsilon_{i}\}_{i=1}^{n} &\stackrel{\mbox{\tiny i.i.d.}}{\sim}
  N(0,\sigma^2). 
\end{align}
Ordinary least squares (OLS) gives a MLE of $ \hat{\bm{\beta}} =
(\mb{X}^\top \mb{X})^{-1} \mb{X}^\top \mb{y}$.  Classically, there are
two main reasons why one may desire a more parsimonious approach to
regression than that provided by OLS.  The first is that OLS tends to
lead to high variance estimators.  The second is a desire for model
fits that have high qualitative interpretability, i.e., that describe
the data adequately but assume no more causes than will account for
the effect.  Our reasons for seeking an alternative are related to the
former more so than the latter.  But, most importantly, we aim to
circumvent the problem of having linear dependence in the columns of
$\mb{Y}_j$ when $n_j \leq j$.  In this case, we are faced with an
$n\times p$ design matrix $\mb{X}$ with number of columns $p$ greater
than the number of observations $n$, yielding an $\mb{X}^\top \mb{X}$
matrix that is singular and cannot be inverted---a so--called ``big
$p$ small $n$'' ($p > n$) problem.  We may even have that $p \gg n$,
say, when the total number of assets $m$ is far greater than the
number of returns recorded for the asset with the shortest history.

Popular solutions to this problem involve methods of variable
selection and coefficient shrinkage.  Probably the most
straightforward method is {\em subset selection} \citep[][Section
3.4.1]{hastie:tibsh:fried:2001} which aims to find the model with the
``best'' size $k$ (i.e., with $k\in \{1,\dots,\min(p,n-1)\}$
covariates).  ``Best'' can be defined in a number of ways, but
typically involves $t-$tests, or minimizing an estimate of expected
prediction error.  Searching through all possible subsets quickly
becomes infeasible for $p>40$.  Larger $p$ can be handled by greedy
methods, but these offer fewer guarantees.  Such methods include {\em
  forward stepwise selection} which starts in the null (intercept
only) model and sequentially adds predictors, and {\em backward
  stepwise selection} which starts at the saturated model (only
applicable when $p<n$) and deletes predictors.  Hybridizations also
exist.

By discarding some predictors, subset selection methods can yield a
model which is more interpretable, and may have lower prediction
error.  But this ``discrete'' process can produce estimators with high
variance. Shrinkage methods are a popular alternative.  They are
hailed for being more ``continuous'', and in some special cases they
can have implicit behavior similar to methods like forward selection.
The following subsection considers the shrinkage methods of ridge
regression, and those related to the lasso.  In Section \ref{sec:pc}
we consider another family of methods which are based on derived input
directions: principal components regression, which has connections to
ridge regression, and partial least squares regression.  These are
handy when the predictors are highly correlated.

The parsimonious regression methods outlined in this section have been
chosen for familiarity, computational tractability, and
implementation.  In each case {\sf R} packages are
available on the Comprehensive {\sf R} Archive Network (CRAN),
\begin{center}
\verb!http://cran.R-project.org! \hspace{1cm} \citep{rproject}, 
\end{center}
\noindent which provide off--the--shelf implementations that will make
for nice subroutines within the framework of constructing estimators
for MVN data under monotone missingness.  It is typical to first
standardize the inputs ($\mb{X}$ and $\mb{y}$) as the methods outlined
below are not equivariant under re-scaling.

\subsection{Shrinkage methods: ridge regression, and the lasso}
\label{sec:ridge}

{\em Ridge regression} and the {\em lasso} shrink the coefficients of
an OLS regression by imposing a penalty on their size:
\begin{equation}
  \hat{\bm{\beta}}^{(q)} = \argmin_{\bm{\beta}}
  \left\{\sum_{i=1}^n \left(y_i - \beta_0 -
      \sum_{j=1}^p x_{ij} \beta_j\right)^2 +
    \lambda \sum_{j=1}^p |\beta_j|^q\right\}
\label{eq:ridge:lasso}
\end{equation}
with $q=2$ for ridge regression, and $q=1$ for the lasso.  The tuning
parameter $\lambda$ controls the amount of shrinkage.  Notice that the
intercept ($\beta_0$) is left out of the penalty term.  Solutions to
(\ref{eq:ridge:lasso}) can be obtained analytically in the case of
ridge regression with $\hat{\bm{\beta}}^{(2)} = (\mb{X}^\top \mb{X} +
\lambda \mb{I})^{-1} \mb{X}^\top \mb{y}$.  Quadratic programming is
required for the lasso.  Both methods have interpretations as Bayesian
{\em maximum a posteriori} (MAP) estimators after imposing particular
prior distributions.  Other choices of $q>0$ are also possible,
however the constraint region for $0<q<1$ is non-convex, which makes
solving the optimization problem more difficult.

For ridge regression, the penalty parameter ($\lambda$) is most
advantageously chosen by minimizing cross validation (CV) estimates of
predictive error.  The commonly used HKB \citep{hkb:1975} and L--W
\citep{lw:1976} methods are computationally efficient, but require
that $p < n$ to fit an OLS.  The implementation of ridge regression
used in this paper comes from the {\tt MASS} library \citep{mass:2002}
for {\sf R} in the form of a function called {\tt lm.ridge}.

Though the form of ridge regression and the lasso are similar, there
are several important differences.  A large $\lambda$ will cause the
ridge estimator $\hat{\bm{\beta}}^{(2)}$ to have many coefficients
shrunk towards zero.  The lasso estimator $\hat{\bm{\beta}}^{(1)}$ has
as similar effect, but, importantly, may contain many coefficients
which are exactly zero---something which is only possible for $0 < q
\leq 1$.  In the Bayesian interpretation, setting $q\leq 1$
corresponds to choosing a prior which concentrates more mass on small
$|\beta_j|$, with the most on $\beta_j = 0$.  In this way, the lasso
implements a kind of continuous subset selection.  As $\lambda$ is
increased, the $|\beta_j|$ decrease, eventually increasing the number
of them which are identically zero, though this relationship need not
be strictly monotonic.

The implementation of lasso used in this paper is contained in the
{\tt lars} package for {\sf R} \citep{lars:2007}.  \cite{efron:2004}
show how the lasso, and two methods called {\em stepwise} and {\em
  forward stagewise}, are special cases of their method of {\em least
  angle regression} (LAR).  LARS can calculate all possible lasso
estimators with computational effort in the same order of magnitude as
OLS regression applied to the full set of covariates.  CV can be used
to select the final model, e.g., using the ``one--standard--error''
rule \citep[][Section 7.10]{hastie:tibsh:fried:2001}, or a more
thrifty $C_p$ \citep{mallows:1973} method can be used, but only when
$p < n$.  When applicable, the $C_p$ method performs nearly as well as
CV within the MVN setting with monotone missingness.
\cite{madigan:ridgeway:2004} come to similar conclusions on equally
tame benchmarks.  However, $C_p$ has also been criticized for
preferring large models \citep{ishwaran:2004,stine:2004} and for being
slightly at odds with LARS \citep{loubes:massart:2004}.  Since we are
mostly interested in applying LARS methods (i.e., lasso) when OLS is
not applicable, i.e., when $p \geq n$, we shall generally rely on CV
to select the final model.

\subsection{Principal components and partial least squares regression}
\label{sec:pc}

In situations where there are a large number of highly correlated
inputs, a decomposition by principal components (PCs) can be used to
select a small number of linear combinations of the original inputs to
be used in place of $\mb{X}$.  The related methods of principal
component regression (PCR) and partial least squares regression (PLSR)
start by performing an orthogonal decomposition of $\mb{X}$, but
differ in how the linear combinations are constructed.

In PCR, {\em singular value decomposition} (SVD) is performed on
$\mb{X}$, i.e., $\mb{X} = (\mb{U} \mb{D}) \mb{V}^\top =
\mb{T}\mb{P}^\top$, where $\mb{U}$ is an $n \times p$ matrix of left
singular vectors describing the ``output basis'', $\mb{D}$ is a
diagonal matrix containing the corresponding singular values (a
square--root of the eigenvalues) in non-decreasing order, $\mb{V}$ is
a $p \times p$ matrix of right singular vectors describing the ``input
basis'', and $\mb{T}$ and $\mb{P}$ are the so--called {\em scores} and
{\em loadings} defined by the decomposition.  Next, $\mb{y}$ is
regressed on the first $k$ PCs, i.e., the scores $\mb{T}_{(k)}$, where
the $(k)$ subscript indicates the extraction of the first $k$ columns
of $\mb{T}$, i.e., the first $k$ columns of $\mb{U}$, $\mb{V}$, and
the first $k$ rows/cols of $\mb{D}$.  Since the columns of $\mb{T}$
are orthogonal, the solution is just a sum of univariate regressions.
Importantly, the solution can then be written in terms of the
coefficients on the predictors in the columns of $\mb{X}$,
\begin{align}
  \mbox{(arbitrary scores and loadings)} && \hat{\bm{\beta}}(k) &=
  \label{eq:preg}
  \mb{P}_{(k)} (\mb{T}_{(k)}^\top \mb{T}_{(k)})^{-1} \mb{T}_{(k)}^\top \mb{y} \\
  \mbox{(from SVD on $\mb{X}$)} && \hat{\bm{\beta}}^{\mbox{\tiny pcr}}(k) &
  =\mb{V}_{(k)} \mb{D}_{(k)}^{-1} \mb{U}_{(k)}^\top \mb{y}, \nonumber
\end{align}
a vector of length $p$.  When $k=p < n$, the coefficients in
(\ref{eq:preg}) are identical to those obtained by OLS. There are many
ways of choosing how many components ($k$) to keep in the final model.
One way is to consider the relative sizes of the eigenvalues as a
proportion of the variation explained by each principal component, and
then choose $k$ so that 80--90\% of the variation is explained.  A
less ad hoc and more reliable---but more computationally
intensive---method that can be applied even when $p \geq n$ involves
using CV to estimate predictive error in order to find $k \in
\{1,\dots,\min(p,n-1)\}$.

PLSR, by contrast, aims to incorporate information about both $\mb{X}$
and $\mb{y}$ in the scores and loadings---which in this context are
often called {\em latent variables} (LVs)---by proceeding iteratively.
The method is initialized with the SVD of $\mb{X}^\top \mb{y}$,
thereby including information about the correlation between, and the
variance within, $\mb{X}$ and $\mb{y}$.  The scores and loadings
obtained by PLSR optimally capture the covariance between $\mb{X}$ and
$\mb{y}$, whereas PCR concentrates only on the variance of $\mb{X}$
\citep{dejong:1993}.  There are several algorithms for obtaining the
scores and loadings, but once obtained, the regression coefficients
$\hat{\bm{\beta}}^{\mbox{\tiny plsr}}(k)$ in $\mb{X}$-space are
recovered by following (\ref{eq:preg}), and CV can be similarly used
to pick $k$.

In situations where a minor component of $\mb{X}$ is highly correlated
with $\mb{y}$, PLSR may have a significant advantage over PCR.
Otherwise, the methods have a more or less comparable performance
record despite a few operational differences---e.g., PLSR usually
needs fewer LVs, but can also yield higher variance estimators of the
regression coefficients.  Both have behavior similar to other
shrinkage methods, particularly ridge regression.  For example, it can
be shown \citep{frank:fried:1993} that ridge regression shrinks the
coefficients of principal components by a factor of
$d_j^2/(d_j^2+\lambda)$, where the $d_j$ are from the diagonal of
$\mb{D}$, whereas PCR truncates them at $k$.

An {\sf R} package called {\tt pls} \citep{heige:2007} provides a
unified implementation of PCR and three algorithms for PLSR
\citep{dayal:macg:1997,dejong:1993,martens:naes:1989}, together with
built--in facilities for estimating $k$ via CV.

\section{The {\tt monomvn} algorithm}
\label{sec:monomvn}

So long as $n_j > j$ for all $j=1\dots,m$, and $n_j \geq n_{j+1}$, an
algorithm for finding the parameters $\bm{\mu}$ and $\bm{\Sigma}$ that
maximize the MVN likelihood for monotone missing data proceeds as
outlined in Section \ref{sec:monotone}.  Initialize $\mu_1$ and
$\Sigma_{11}$ to the sample mean and variance of the first column
$\mb{y}_1$ of $\mb{Y}$, then iterate through the following steps for
$j=2,\dots,m$:
\begin{enumerate}
\item Find the MLEs (\ref{eq:regress}) of $\bm{\beta}_j$ and
  $\sigma_j^2$ in a regression (\ref{eq:monoreg}) of $\mb{y}_j$ onto the first
  $j-1$ columns of $\mb{Y}$ (as predictors), using only the first
  $n_j$ observations;
\item Obtain the MLEs of $\mu_j$ and $\mb{\Sigma}_{(1:j),j}$
  from $\hat{\bm{\mu}}_{1:(j-1)}$, $\hat{\bm{\Sigma}}_{1:(j-1),1:(j-1)}$,
  $\hat{\bm{\beta}}_j$ and $\hat{\sigma}^2_j$ as in (\ref{eq:addy}).
\end{enumerate}
If any $n_j \leq j$, then we have a ``big $p$ small $n$'' problem, and
the standard regression in step 1 above cannot be performed.  In
practice, it may be that $n_j > j$ and still there are columns of the
design matrix which are not linearly independent, and so it is not of
full rank.  The chances that this may happen become increasingly more
likely as $j$ approaches $n_j$ when finite (double--precision)
computer representations make it so that the design matrix is
numerically rank deficient.  Both issues are addressed simultaneously
by instead performing one of the parsimonious regressions outlined in
Section \ref{sec:bpsn}.  Then step 2 can proceed as usual.  Observe
that this approach also enables estimation when there are more assets
than historical returns ($m > n$).

\subsection{Choosing the parsimonious proportion}

Even when parsimonious regression is not strictly necessary, it can
aid in interpretation, and possibly even yield more accurate and lower
variance estimators.  The lasso and the other LARS methods can
choose to shrink $\bm{\beta}$ so that only the intercept term is
nonzero.  This enables the detection of zeros in the MVN covariance
matrix $\bm{\Sigma}$.  In other words, it can be used as a test, of
sorts, for independence between assets.  

Towards building a more efficient and interpretable estimator, one may
consider applying a parsimonious regression for every iteration of
step 1 above.  This is explored further in Section \ref{sec:depend}.
Alternatively, one could determine a threshold, say $p$, representing
a proportion of rows to columns in the design matrix past which a
parsimonious regression is applied regardless.  That is, when $n_j
\leq pj$, for $0\leq p\leq 1$.  Then, the $p=0$ case corresponds to
always using a parsimonious method, and $p=1$ reverts to applying one
only when necessary.  In Section \ref{sec:parsi} we show how easy it
is to establish reliable rules of thumb for choosing $p$.

\subsection{Incorporating factors}
\label{sec:fact}

A popular estimator for the covariance matrix of financial asset
returns involves using {\it factor models}.  The essential idea behind
the factor model is to regress the observed returns $\mb{y}_j$ on
measured common market factors $\mb{F}$, and to derive a covariance
matrix of the returns as a function of the regression equations.

For a factor space with $K$ factors, the model can be formalized as
follows. Each excess return $y_{i, j}$ is modeled by the regression
equation
\begin{equation}
\label{eq:factor-regression}
y_{i, j} = \lambda_{0, j} + \sum_{k = 1}^K \lambda_{k, j}f_{i, k} + \epsilon_{i, j}
\end{equation}
where each $\epsilon_{i, j}$ is a residual term independent of
$\mb{F}$. The residual terms for the $i^{\mbox{\tiny th}}$ instance
are assumed to follow a zero--mean MVN with diagonal covariance matrix
$\mb{D}$.  For instance, a common one--factor model takes $f$ to be
value--weighted market index \citep[e.g.,][]{ckl:1999}. A common
three--factor model augments the value--weighted market index with
size and book--to--market factors \citep{famafrench:1993}.

Factors are assumed, for now, to be i.i.d.~and to follow a MVN with
$K\times K$ covariance matrix $\bm \Omega$.  Let $\bm{\Lambda}$ be the
$K\times m$ matrix defined by the entries $\bm{\Lambda}_{k, j} =
\lambda_{k, j}$, for $k=1,\dots,K$.  It follows that the covariance
matrix of the returns, as parameterized by $\{\bm{\Omega},
\bm{\Lambda}, \mb{D}\}$, is given by
\begin{equation}
\bm{\Sigma}^{(f)} = \bm{\Lambda}^\top \bm{\Omega}\bm{\Lambda} + \mb{D}.
\end{equation}
An estimate $\hat{\bm{\Sigma}}^{(f)}$ can therefore be obtained by
estimating each column $\hat{\bm{\lambda}}_j = (\lambda_{1, j},
\dots,\lambda_{K, j})^\top$ of $\hat{\bm{\Lambda}}$ by regressing
$\bm{y}_j$ on $\mb{F}$ with an intercept.  The mean sum of squares of
the residuals of each regression forms the diagonal of $\hat{\mb{D}}$,
and the off--diagonal entries are zero.  The estimate $\hat{\bm
  \Omega}$ is the empirical covariance of the factors.  Note that each
regression equation requires only the data observed for the particular
return $\mb{y}_j$, together with the corresponding observations for
the factor(s).  However in practice, the method is applied only to
completely observed $\mb{Y}$ and $\mb{F}$.

The main underlying assumption is that returns are mutually
independent conditioned on the factors. If the number of factors is
considerably smaller than the number of returns, the model will be
parsimonious and the resulting $\hat{\bm{\Sigma}}^{(f)}$ will have
lower variance than the empirical covariance matrix.  This assumption
allows for any missingness pattern, even the extreme one where no
joint observation of returns $\mb{y}_j$ and $\mb{y}_k$ exists.  The
drawback is that the independence assumptions encoded in this model
might be unrealistic, and the resulting estimate will suffer from a
strong bias.

Instead, we can use the data to find which independence assumptions
are adequate by integrating the factor model into the {\tt monomvn}
framework.  Consider the {\it full} regression model, where we regress
$\mb{y}_j$ on $\mb{Y}_j$ and $\mb{F}_j \equiv\mb{F}_{1:(j-1)}^{(n_j)}$
simultaneously:
\begin{equation}
\label{eq:full-factor-regression}
\mb{y}_j = \mb{Y}_j \bm{\beta}_j + \mb{F}_j
\bm{\lambda}_j + \bm{\epsilon}_j,
\end{equation}
The $\lambda_{0, j}$ term does not appear because it is not
identifiable given the presence of $\beta_{0, j}$.  Since this
formulation is in the same family of parameterizations of the original
models used in {\tt monomvn}, an analogous procedure applies with
minor pre- and post-processing. First shift the labels the returns for
each asset by $K$ so that $\mb{y}_j$ becomes $\mb{y}_{j + K}$ and the
corresponding $\bm{\beta}_j$ becomes $\bm{\beta}_{j+K}$.  Then map
$\mb{F}_k$ to $\mb{Y}_k$ and $\bm{\lambda}_k$ to $\bm{\beta}_k$.  If
the recursion in Eq.~(\ref{eq:addy}) is then applied as usual, giving
the estimates $\hat{\bm{\mu}}$ [an $(m+K)$ vector] and $\hat{\bm{\Sigma}}$
[an $(m+K)\times (m+K)$ matrix], an estimate of the covariance matrix
of the asset returns can then be extracted from the bottom--right $m
\times m$ block of $\hat{\bm{\Sigma}}$, i.e.,
$\hat{\bm{\Sigma}}^{(f+m)} = \hat{\bm{\Sigma}}_{(K + 1):(m + K), (K +
  1):(m + K)}$.  The superscript $(f+m)$ is meant to indicate
dependence on both factors and assets.  Importantly, no internal
changes to the workings of the {\tt monomvn} algorithm are necessary.

Observe that if the (parsimonious) regression method applied within
{\tt monomvn} uses OLS whenever regressing onto the factors, and sets
the regression coefficients to zero otherwise, then we obtain
$\hat{\bm{\Sigma}}^{(f+m)} = \hat{\bm{\Sigma}}^{(f)}$.  In the context
of {\tt monomvn} we call this the ``factor--parsimony'' regression,
filling a role similar to PCR, lasso, etc.  If required, the
covariance matrix of the factors can also be recovered as
$\hat{\bm{\Omega}} = \hat{\bm{\Sigma}}_{1:K,1:K}$.  Also observe that,
within the {\tt monomvn} framework, it is possible to handle factors
with historical missingness.

If, instead of the factor--parsimony method, any of the other methods
(outlined in Section \ref{sec:bpsn}) are used, then shrinkage is
applied to both $\bm \beta_j$ and $\bm{\lambda}_j$ in
(\ref{eq:full-factor-regression}).  In this case we obtain a
generalization of the independence structure assumed in the classical
factor model, allowing the data (factors and returns) to determine the
appropriate mix of influence on the resulting estimator for
$\bm{\Sigma}$.  It is interesting to point out the link between this
generalized factor model (\ref{eq:full-factor-regression}) resulting
in $\hat{\bm{\Sigma}}^{(f+m)}$, and the optimal shrinkage estimator of
\citet{ledoit:2002}:
\begin{equation}
  \hat{\bm{\Sigma}}^{(\ell)} = \alpha \hat{\bm{\Sigma}}^{(f)} + 
  (1 - \alpha)\hat{\bm{\Sigma}}^{(c)}, \;\;\;\;\; \mbox{for } \alpha \in [0, 1].
  \label{eq:ledoit}
\end{equation}
Here, $\hat{\bm{\Sigma}^{(c)}}$ is the standard covariance estimate
obtained using only the portion of the data available across all
assets and $\alpha$ is an ``optimal'' mixing proportion chosen by CV.
(Note that Ledoit's factor--based estimator $\hat{\bm{\Sigma}}^{(f)}$
uses only completely observed joint returns.)  The spirit of these two
approaches is similar, but they are quite distinct.  The published
success of this type of shrinkage approach suggests that it is
important to combine a (complete data) factor--based estimate with a
traditional covariance estimate.  Indeed, the estimator
$\hat{\bm{\Sigma}}^{(f+m)}$ involves combining covariances mediated by
factors with covariances that are not accounted for by factors; it can
also handle historical missingness via the ``factor--parsimony''
regressions within {\tt monomvn}.  But rather than shrinking a
(possibly) non--positive definite estimator $\hat{\bm{\Sigma}^{(c)}}$
towards $\hat{\bm{\Sigma}^{(f)}}$ with a single parameter $\alpha$ as
in (\ref{eq:ledoit}), {\tt monomvn} applies $m+K$ unique shrinkage
parameters, one for {\em each} regression, while taking full advantage
of all available returns.

\subsection{Software}

Finally, an {\sf R} package called {\tt monomvn} \citep{monomvn} has
been made freely available through CRAN. It implements the algorithm
described in this section, and supports all of the parsimonious
regression methods outlined in Section \ref{sec:bpsn} via the
stand--alone packages outlined therein.  Two forms of CV are supported
for choosing the number of components in the parsimonious regression:
random 10--fold and (deterministic) leave--one--out (LOO).  A $p$
argument facilitates parsimonious regression modeling, as described
above.  Incorporating factors is as straightforward as bundling them in
as if they were returns, as described above.

\section{Empirical results}
\label{sec:results}

In this section, the {\tt monomvn} methods are illustrated and
validated on real and synthetic data.  In Section \ref{sec:synth} we
focus on the properties of estimates of $\hat{\bm{\mu}}$ and
$\hat{\bm{\Sigma}}$ in a controlled setting involving synthetic data
under monotone missingness.  In \ref{sec:portfolio} we turn to
applying the estimators towards balancing portfolios in a
mean--variance setting.  We wrap up in \ref{sec:depend} by using
{\tt monomvn} in a descriptive analysis of dependence involving
thousands of assets.

\subsection{Properties of the estimators on synthetic data}
\label{sec:synth}

Here, we use a data--generation mechanism provided by the {\tt
  monomvn} package: {\tt randmvn} generates random samples from a
randomly generated MVN distribution with an i.i.d.~standard normal
mean vector $\bm{\mu}$, and an Inv--Wishart sampled $\bm{\Sigma}$;
{\tt rmono} imposes a uniformly distributed monotone missingness
pattern.  A similar method is used to generate samples with monotone
missingness from a multivariate $t$ distribution (MV$t$) as well, in
order to demonstrate that the MVN--based {\tt monomvn} methods still
perform well in the presence of heavier tailed data.


The comparisons to follow focus on highlighting the relative strengths
and weaknesses of variations of {\tt monomvn} as a function of the
choice of parsimonious regression method applied.  Additionally, two
simpler methods are devised as calibration tools, and to illustrate
the advantage of the {\tt monomvn} approach over those which do not
leverage the structure of the monotone missingness pattern.  The
simplest comparator is called ``complete'', where $\bm{\mu}$ and
$\bm{\Sigma}$ are estimated using only the portion of data available
across all assets, i.e., only the completely observed returns.  Put
yet another way: only the first $n_m$ rows of $\mb{Y}$ are used.
Another comparator is ``observed'' which uses all of the available
data in an obvious but na\"ive way:
\begin{align}
  \hat{\mu}_j &= \frac{1}{n_j} \sum_{k=1}^{n_j} y_{k,j} && \mbox{
    and} & \hat{\Sigma}_{i,j} &= \frac{1}{n_j} \sum_{k=1}^{n_j}
  (y_{k,j} - \hat{\mu}_j)(y_{k,i} - \hat{\mu}_i) \;\;\;\; \mbox{ for }
  i=1,\dots,j.
\end{align}
Unfortunately, the covariance matrices provided by the ``complete'' and
``observed'' estimators are not guaranteed to be positive--definite
\citep{stambaugh:1997}.  

As a final comparator, we consider a method of estimation for
incomplete data for arbitrary missingness patterns
\citep{dempster:laird:rubin:1977}, using the expectation conditional
maximization (ECM) algorithm \citep{meng:rubin:1993}.  Consequently,
this method also works when the missingness pattern is monotone, but
represents a sort of overkill in this case.  Two similar software
packages are available for this method when the data is assumed to
follow a multivariate normal distribution: the {\tt norm} package
\citep{norm:2002} for {\sf R}, and {\tt ecmnmle} (contained in the
{\sf Matlab} {\tt Financial Toolbox}).  We prefer {\tt norm} because
its core is implemented in compiled {\sf Fortran}, with an {\sf R}
wrapper.  It gives nearly identical results to---but runs more than 20
times faster than---{\tt ecmnmle} which is written solely in {\sf
  Matlab}.  The ECM method iterates until convergence, stopping at a
{\em local} maximum when an improvement threshold is met.  As a
result, its computational demands and the ultimate optimality of the
resulting estimator are sensitive to the initial configuration of the
algorithm.  Though the missingness pattern may be arbitrary, it is
well--known that the method can fail due to convergence issues and/or
numerical singularities that can arise due to finite machine
representations when more than 15\% of the data is missing (see, e.g.,
the {\tt ecmnmle} documentation within {\sf Matlab}).  So it cannot
handle $m > n$, which precludes it from general use in our problem.

The expected log likelihood (ELL), which is related to the
Kullback--Leibler (KL) divergence, is used as the main metric for
comparisons.  For probability distribution functions (PDFs) $p$ and
$q$, the KL divergence between $p$ and $q$ is defined as
\[
D_{\mbox{\tiny KL}}(q \parallel p) = \int p(x) \log \frac{p(x)}{q(x)} \;dx.
\]
In the particular case where $q$ is the estimated MVN with parameters
$\hat{\bm{\mu}}$ and $\hat{\bm{\Sigma}}$ and $p$ is the ``true''
parameterization with $\bm{\mu}$ and $\bm{\Sigma}$, the KL divergence
can be shown to be:
\[
D_{\mbox{\tiny KL}}(\mathrm{MVN}(\hat{\bm{\mu}}, \hat{\bm{\Sigma}}) \parallel
\mathrm{MVN}(\bm{\mu}, \bm{\Sigma})) = \frac{1}{2} \left(\log
  \frac{|\hat{\bm{\Sigma}}|}{|\bm{\Sigma}|} +
  \mbox{tr}(\hat{\bm{\Sigma}}^{-1} \bm{\Sigma}) + (\hat{\bm{\mu}} -
  \bm{\mu})^\top \hat{\bm{\Sigma}}^{-1}(\hat{\bm{\mu}} - \bm{\mu}) \right).
\]
The ELL of $q$ relative to data sampled from $p$ is given by
\begin{align}
\mathbb{E}_p\{\log q\} &= \int p(x) \log q(x) \;dx \nonumber \\
&= \int p(x) \log p(x) \;dx
- D_{\mbox{\tiny KL}}(q \parallel p). \label{e:ell}
\end{align}
The integral $\int p\log p$ in (\ref{e:ell}) is the entropy of $p$.
The entropy of $\mathrm{MVN}(\bm{\mu}, \bm{\Sigma})$ can be shown to
work out to $-\frac{1}{2} \log \{(2\pi e)^N |\bm{\Sigma}|\}.  $ When
analytical expressions are not available it is easy to approximate
(\ref{e:ell}) numerically by $T^{-1} \sum_{t=1}^T \log q(x_t)$, where
$x_t \sim p$ is simulated out of sample.  This nicely converges to the
truth for large $T$.  The ELL is good for ranking competing
estimators, however actual ``distances'' between estimators is hard to
interpret.


\subsubsection{Comparing estimators}

\begin{figure}[ht!]
\centering
\includegraphics[angle=-90, scale=0.285]{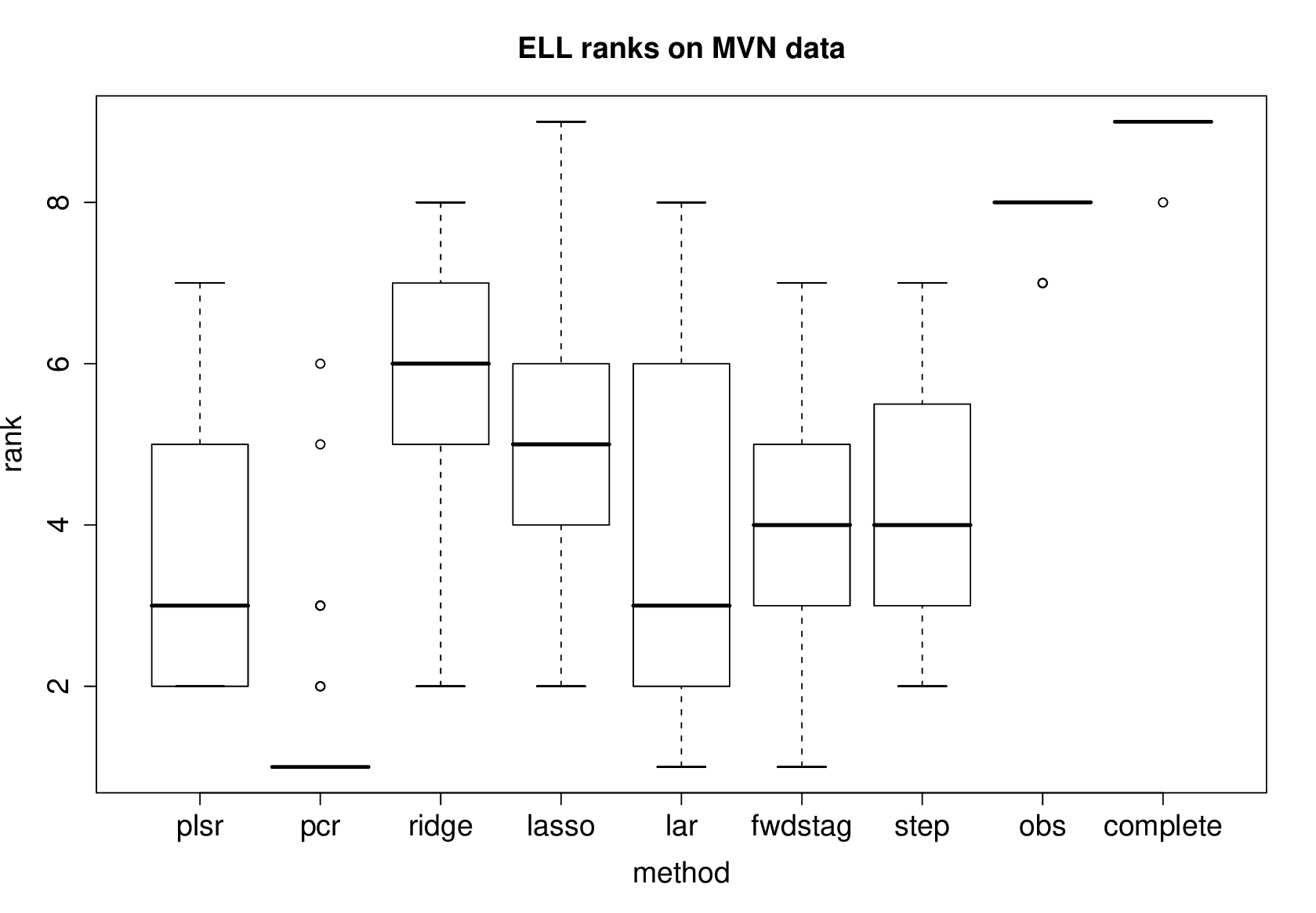}
\includegraphics[angle=-90, scale=0.285]{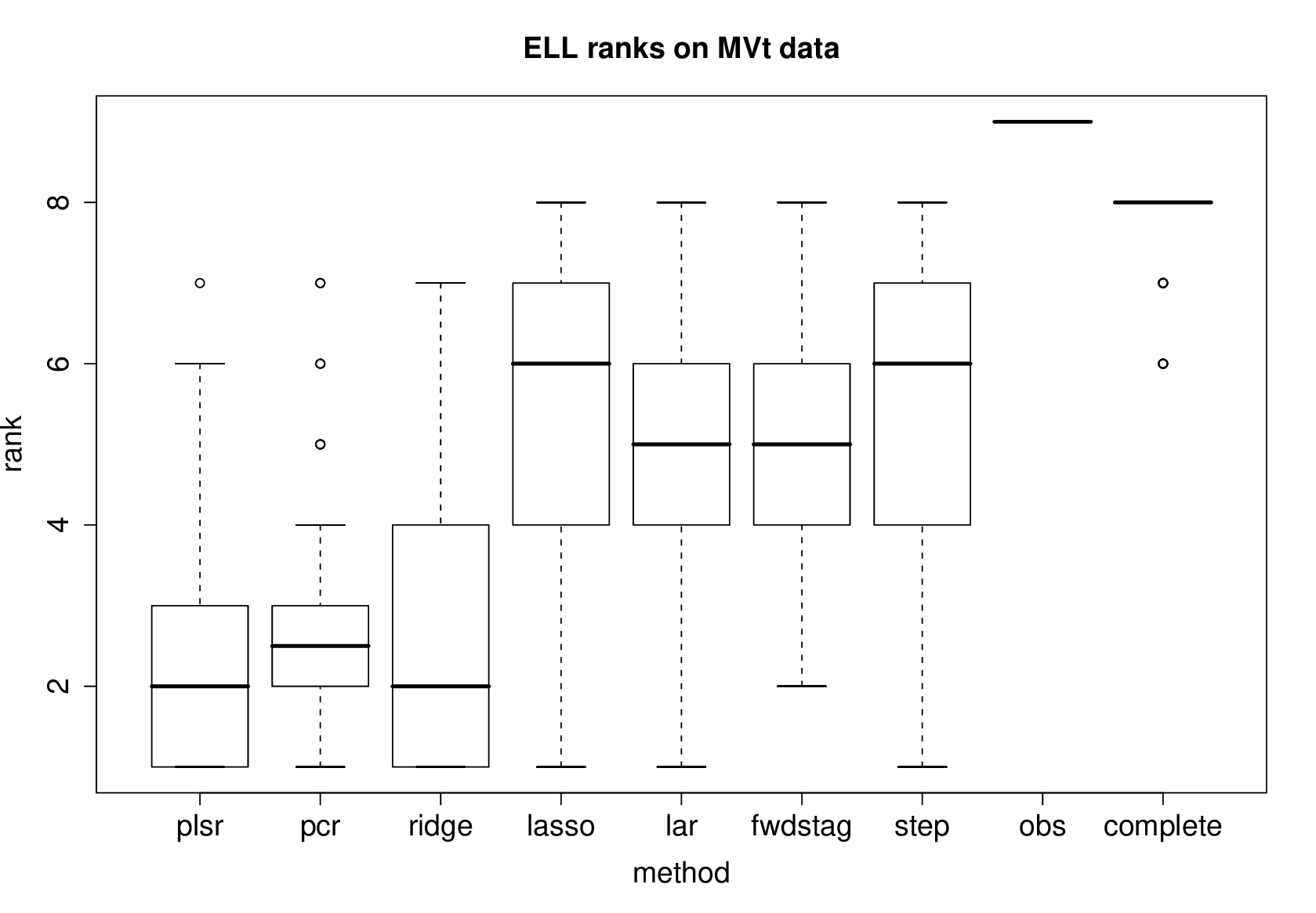}
\caption{Comparison of parsimonious regression ($p=1$) methods (using
  10--fold CV) on randomly generated MVN data ($n=1000$ samples,
  $m=100$ dimensions) data with $\bm{\mu}\sim N_m(0,1)$, $\bm{\Sigma}
  \sim$ Inv--Wishart and uniform monotone missingness: boxplots of ELL
  ranks summarizing 100 repeated trials.
  \label{f:synth}}
\end{figure}

Figure \ref{f:synth} {\em (left)} summarizes a comparison between the
different parsimonious regressions within the {\tt monomvn} algorithm,
using randomly generated MVN data with $m=100$ and $n=1000$, repeated
over 100 trials, each time sampling new $\bm{\mu}$, $\bm{\Sigma}$ and
$\mb{Y}\sim \mathrm{MVN}(\bm{\mu}, \bm{\Sigma})$ with uniform monotone
missingness.
Parsimonious regressions were used only when necessary (i.e., $p=1$).
10--fold CV was used to choose $\lambda$ or the number of (principal)
components.  As can be seen from the table, PCR emerges as the clear
winner in this comparison, nearly always having the best ELL rank.
The complete and observed comparators are almost always ranked worst.

In anticipation of the application in Section \ref{sec:portfolio} to
financial returns data, which are believed to follow a heavier tailed
distribution than MVN, we repeated the above experiment with
synthetically generated MV$t$ data with a monotone missingness
pattern.  The degrees of freedom parameter was sampled as $\nu \sim
\mathrm{Exp}(\frac{1}{2})+1$.  Figure \ref{f:synth} {\em (right)} shows
roughly similar behavior for the MVN based {\tt monomvn} estimators
when fit to MV$t$ data: PCR is the best and the observed and complete
estimators are the worst (although the order is switched).  ELL was
computed numerically using the known degrees of freedom parameter(s),
$\nu$, which generated the data.  This is a legitimate choice since
the $\nu$ is not used in the mean--variance analysis to follow in
Section \ref{sec:portfolio}.  It is interesting to note the improved
rank(s) of the ridge regression based estimator in this case.

These results are in line with those of previous simulation studies
which compare ML estimators---that are able to leverage all of the
available data by exploiting the MVN assumption---to those which use
more reasonable distributional assumptions but which, for reasons of
tractability, can only use the completely observed cases
\citep[e.g.,][]{little:1988}.  The evidence suggests that making use
of all of the available data in a sensible way is the crucial
ingredient despite that the underlying assumptions may be violated.
The dominance of PCR in both MVN and MV$t$ scenarios is in line with a
recent study \citep{cpr:5829} showing that PCR out--competes other
shrinkage (Bayesian motivated) estimators in applications with a large
number of financial asset returns.

\subsubsection{Choosing the parsimonious proportion}
\label{sec:parsi}

Recall from Section \ref{sec:monomvn} that $p\in [0,1]$ determines
when a parsimonious method is to be used instead of OLS in the {\tt
  monomvn} algorithm.  The experiment performed here is similar to the
previous one, except that $n$ and $m$ are varied stochastically with
$m$ uniform in $\{5,\dots,100\}$ and $n|m$ uniform in $\{\max(10,
\lfloor m/2\rfloor),\dots, md\}$.
\begin{table}[ht]
\begin{center}
\begin{tabular}{l||rrr|r}
& \multicolumn{3}{c|}{optimal $p$} & \\
method & 5\% & mean & 95\% & improv \\
  \hline
  plsr & 0.12 & 0.23 & 0.37 & 0.55  \\
  pcr & 0.09 & 0.27 & 0.51 & 0.69 \\
  ridge & 0.04 & 0.25 & 0.67 & 0.29 \\
  lasso & 0.12 & 0.24 & 0.38 &  0.76 \\
  lar & 0.11 & 0.26 & 0.41 & 0.65 \\
  stepwise & 0.15 & 0.26 & 0.39 & 0.74
\end{tabular}
\end{center}
\caption{Mean and 90\% interval for optimal $p$, the ratio of columns
  to rows in the design matrix before switching from OLS to a parsimonious
  regression.  The {\em improv} column gives the proportion of runs for
  which $p=0.25$ is better than $p=0$. We repeated this over 100 trials 
  with LOO CV with the ELL as an objective.
\label{t:p}}
\end{table}
Table \ref{t:p} shows the mean and 90\% interval for the optimal $p$
over 100 repeated trials sampling new $m$, $n$, etc., each time.  LOO
CV was used to choose $\lambda$, or the number of (principal)
components, and the objective criteria used was ELL. The final column
in the table shows the proportion of time when $p=0.25$ was better
than $p=0$.  Observe that all methods except ridge regression work
well, as a rule of thumb, with $p=0.25$.  All things being equal, a
larger $p$ setting may be preferred for speed reasons.

\subsubsection{Comparing to ECM}

Due to the limitations of ECM--based methods, like those implemented
by {\tt norm} and {\tt ecmnmle}, a comparison of {\tt monomvn} to
these approaches requires a more controlled experiment.  Fixing $m=10$
and $n=100$, 1000 repeated experiments similar to the ones described
above, with uniform monotone missingness, gave that {\tt monomvn}
(with PCR) had higher ELL 997 times ($100\%$) and that ECM failed to
converge 53 times ($\approx 5\%$).  As $n$ grows relative to $m$, the
performance of the methods converge.  For example, with $m=10$ and
$n=1000$ the means are {\tt monomvn} is better 831 times ($83\%$), and
ECM failed to converge 11 times ($1\%$).  As the dimensionality ($m$)
increases modestly compared to the sample size ($n$), the ECM--based
{\tt norm} algorithm consistently diverges.  For example, with $m=20$
and $n=100$ {\tt norm} fails to converge more than 40\% of the time.

\subsection{Constructing portfolios from historical returns}
\label{sec:portfolio}

In this section we examine the characteristics of minimum variance
portfolios constructed using estimates of $\mb{\Sigma}$ based on
historical monthly returns.  The experimental setup is similar to ones
that have been used in several recent papers on covariance estimation,
and minimum variance portfolio balancing
\citep[e.g.][]{ckl:1999,jagma:2003}.  Following these works we use the
monthly returns of common domestic stocks traded on the NYSE and the
AMEX from April 1968 until 1998. We require that the stocks have a
share price greater than \$5 and a market capitalization greater than
20\% based on the size distribution of NYSE firms.  Estimators of
$\bm{\Sigma}$ are constructed based on (at most) the most recently
available 60 months of historical returns.  This is in keeping with
previous work and acknowledges that the i.i.d.~assumption in
Eq.~(\ref{eq:iidlik}) is only valid locally (in time) due to the
conditional heteroskedastic nature of financial returns.  Short
selling is not allowed; all portfolio weights must be nonnegative.
Although it is typical to cap the weights as well, e.g., at 2\%, in
order to ``tame occasional bold forecasts'' \citep{ckl:1999} that
typically arise due to poor estimators \citep{jagma:2003}, we
specifically do not do so here.  Our goal is fully expose the quality
of the estimators and to illustrate that with good estimators such
rules of thumb are unnecessary.

Four classes of estimators of $\bm{\Sigma}$ are used in the
comparisons which follow.  (1) The {\em complete} estimator outlined
earlier, with variations depending on how many assets have historical
returns with certain lengths (more below). (2) A one--factor model
using the return on the value--weighted portfolio of stocks traded on
the NYSE, AMEX, and Nasdaq. (3) The {\tt monomvn} method using the
parsimonious regressions of Section \ref{sec:bpsn} with $p=0.25$. (4)
The {\tt monomvn} method incorporating the value--weighted portfolio
as a factor with, as described in Section \ref{sec:fact}, and with
$p=0$.  For this class we augment the collection of parsimonious
regressions to include the ``factor--parsimony'' method.  We do not
compare to the ECM methods of {\tt norm} or {\tt ecmnmle} here, as
this has proved to be both cumbersome and troublesome; the methods
seem unable to handle the missingness level in this data.  For
example, {\tt norm} consistently fails to converge even after
thousands of very slow iterations of ECM (each taking several seconds
on a 3.2 GHz Xeon).

To assess the quality and characteristics of the constructed
portfolios we follow \cite{ckl:1999} in using the following:
(annualized) return and standard deviation; (annualized) Sharpe ratio
(average return in excess of the Treasury bill rate divided by the
standard deviation); (annualized) tracking error (standard deviation
of the portfolio return in excess of the S\&P500 return); correlation
to the market (S\&P500 return); average number of stocks with weights
above 0.5\%. We closely follow the experimental setup of
\citet{ckl:1999} and \citet{jagma:2003} by randomly subsampling from
the qualifying stocks in each year, and holding the portfolios for the
entire subsequent 12 months.  The random subsample reduces the size of
the estimation problem, and thus computational burden, so that many
methods can be simultaneously benchmarked against one another.  It can
also serve the dual purpose of enabling the calculation of
nonparametric (bootstrap--like) Monte Carlo assessments of
variability, which was not a feature explored in previous work.

Specifically, in each April, starting in 1972, we randomly subsample
250 stocks
(without replacement) from those which qualify (in the
sense outlined above) and which have at least 12 months of historical
returns.  In this way our work differs slightly from our predecessors
whose estimators require exactly 60 months of historical returns.  We
chose 12 months in order to highlight the benefit of incorporating
assets in the portfolio with fewer than 60 months of returns via {\tt
  monomvn}.  Estimates of the covariance matrix of monthly excess
returns (over the monthly Treasury Bill rate) are generated form the
different models using at most the last 60 months of historical
returns for the 250 assets.  Based on the estimate(s), quadratic
programming is used to find the global minimum variance portfolio(s)
described by weights $\hat{\mb{w}} = \mbox{argmin}_{\mb{w}} \mb{w}^T
\mb{\hat{\bm{\Sigma}}} \mb{w}$.  Then, the weights $\hat{\mb{w}}$ are
applied to form buy--and--hold portfolio returns until the next April,
when the randomization, estimation, and optimization steps are
repeated and the portfolios are reformed.

\begin{table}[ht!]
\begin{center}
\begin{tabular}{r||rrrrrr}
method & mean & sd & sharpe & te & cm & wmin \\
  \hline \hline
eq & 0.149 & 0.188 & 0.432 & 0.062 & 0.949 & 0 \\
vw & 0.135 & 0.162 & 0.412 & 0.032 & 0.981 & 45 \\
\hline
min & 0.147 & 0.183 & 0.431 & 0.105 & 0.819 & 29 \\
com & 0.150 & 0.183 & 0.447 & 0.107 & 0.810 & 26 \\
rm & 0.132 & 0.129 & 0.494 & 0.094 & 0.803 & 16 \\
\hline
fmin & 0.142 & 0.146 & 0.503 & 0.086 & 0.845 & 38 \\
fcom & 0.144 & 0.146 & 0.521 & 0.087 & 0.841 & 37 \\
frm & 0.138 & 0.130 & 0.537 & 0.117 & 0.688 & 21 \\
\hline
plsr & 0.148 & 0.154 & 0.516 & 0.124 & 0.686 & 15 \\
pcr & 0.143 & 0.132 & 0.563 & 0.109 & 0.732 & 23 \\
ridge & 0.158 & 0.165 & 0.546 & 0.122 & 0.716 & 16 \\
lasso & 0.151 & 0.150 & 0.550 & 0.054 & 0.941 & 69 \\
lar & 0.151 & 0.151 & 0.545 & 0.053 & 0.944 & 71 \\
step & 0.152 & 0.155 & 0.541 & 0.052 & 0.946 & 75 \\
\hline
ffp & 0.143 & 0.132 & 0.566 & 0.113 & 0.712 & 24 \\
fplsr & 0.147 & 0.153 & 0.514 & 0.123 & 0.688 & 15 \\
fpcr & 0.142 & 0.131 & 0.560 & 0.109 & 0.732 & 24 \\
fridge & 0.158 & 0.163 & 0.554 & 0.119 & 0.726 & 19 \\
flasso & 0.152 & 0.148 & 0.561 & 0.056 & 0.936 & 69 \\
flar & 0.151 & 0.151 & 0.546 & 0.053 & 0.943 & 70 \\
fstep & 0.154 & 0.153 & 0.558 & 0.055 & 0.939 & 73 \\
\end{tabular}
\end{center}
\caption{Comparing statistics summarizing the returns of
  yearly buy--and--hold portfolios generated over 50 repeated 
  random paths through the 26 years of monthly historical returns.
  The first group of rows show the equal-- and value--weighted
  portfolios; the second group of rows have complete data estimators 
  based on the preceding 12--months of returns, the maximal completely
  observed historical returns, and the returns for the subset of
  assets with 60 months of historical returns; the third group 
  uses the same returns as the second with a one--factor model;
  the penultimate group uses {\tt monomvn}; the final group uses
  {\tt monomvn} with the additional one--factor.  The statistics
  across the columns are (annualized) mean return, standard
  deviation, Sharpe ratio, tracking error, correlation to market
  and average number of stocks with weights above 0.5\%.
} \label{t:sharpe}
\end{table}

Table \ref{t:sharpe} summarizes the properties of those returns
averaged over 50 repeated random paths through the 26 years in the
study.  The table is broken into five sections, vertically, starting
with the equal-- and value--weighted portfolios (for comparison),
followed by global minimum variance portfolios based on estimated
$\bm{\Sigma}$: complete data estimators, complete data estimators
based on a one--factor model, {\tt monomvn} estimators, and {\tt
  monomvn} estimators incorporating the one--factor.  Throughout, the
``f'' prefix indicates that the estimator uses the value--weighted
factor in some way.  The ``min'' and ``fmin'' estimators use only the
last 12--months of historical returns, whereas the ``com'' and
``fcom'' estimators use the maximal complete history available.  The
``rm'' and ``frm'' estimators focus only on those assets with
completely observed returns for the last 60 months---where the weights
for the other assets are set to zero (removing them from the
portfolio). The annualized mean, standard deviation, and Sharpe ratio
statistics for these six estimators lead one to conclude that the more
historical returns (within the five--year window) that can be used to
estimate $\bm{\Sigma}$ the better.  Tracking error is also improved,
except in the case of ``frm''.  All in all, these results support
those obtained in previous studies \citep[e.g.,][]{ckl:1999} showing
that, in particular, factor models improve upon the na\"ive estimator
in the complete data case.  Further inspection of this part of the
table reveals that the improved Sharpe ratios for ``rm'' and ``frm''
are due to the smaller standard deviation obtained under these
estimators, but that this comes at the expense of a smaller mean
return.  This may be due to more weight being placed on fewer assets
(as indicated in the ``wmin'' column).  Both ``rm'' and ``frm'' also
have the lowest correlation to the market in their cohort.

The final two groups of rows tell a similar story.  The Sharpe ratios
for the {\tt monomvn} estimators---with and without the
value--weighted factor---show marked improvements over the complete
data estimators.  As before, the inclusion of the value--weighted
factor further adds to the improvement, e.g., yielding higher Sharpe
ratios except in the case of PCR where they remain essentially
unchanged.  The ``ffp'' estimator, i.e., the one--factor model
applied via {\tt monomvn} using the ``factor--parsimony'' regression
method, has the lowest standard deviation, and therefore a
comparatively high Sharpe ratio despite a low mean return.  We can see
that, as with ``rm'' and ``frm'', this low standard deviation is
obtained by placing large weight on only a few assets.  PCR, PLSR, and
ridge regression---both with and without factors---show similar
properties.  In contrast, the LARS estimators (lasso, lar, and
stepwise---both with and without the factor), obtained similar or
better Sharpe ratios but with a large mean return, by assigning large
weight to roughly three times more assets.  As a result, these LARS
estimators obtain a much lower tracking error and higher correlation
to the market.

So when appropriate factors are available it makes sense to use them,
and the best way to do so is via {\tt monomvn}.  It would seem that
the one--factor LARS based {\tt monomvn} estimators give the best
results in the study, overall, with lasso in the top spot.  It is
reassuring to notice that, when an appropriate factor is {\em not}
available, the LARS based {\tt monomvn} methods, and PCR, give largely
similar results by incorporating all of the available returns in a
parsimonious way.  This is not true in the case of the complete data
estimators.

\begin{figure}[ht!]
\centering
\includegraphics[trim=40 0 0 10,scale=0.75]{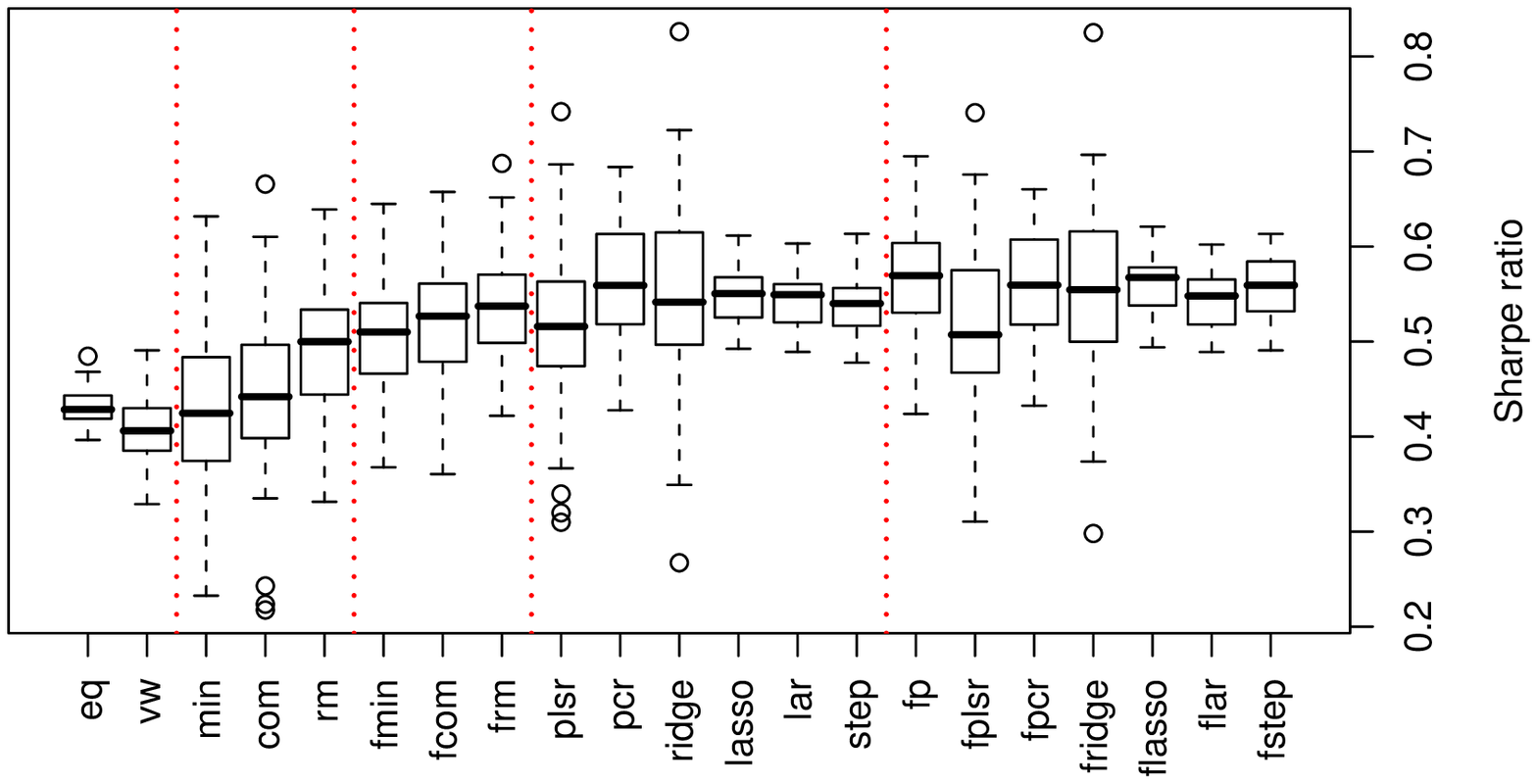}
\includegraphics[trim=40 0 0 25,scale=0.75]{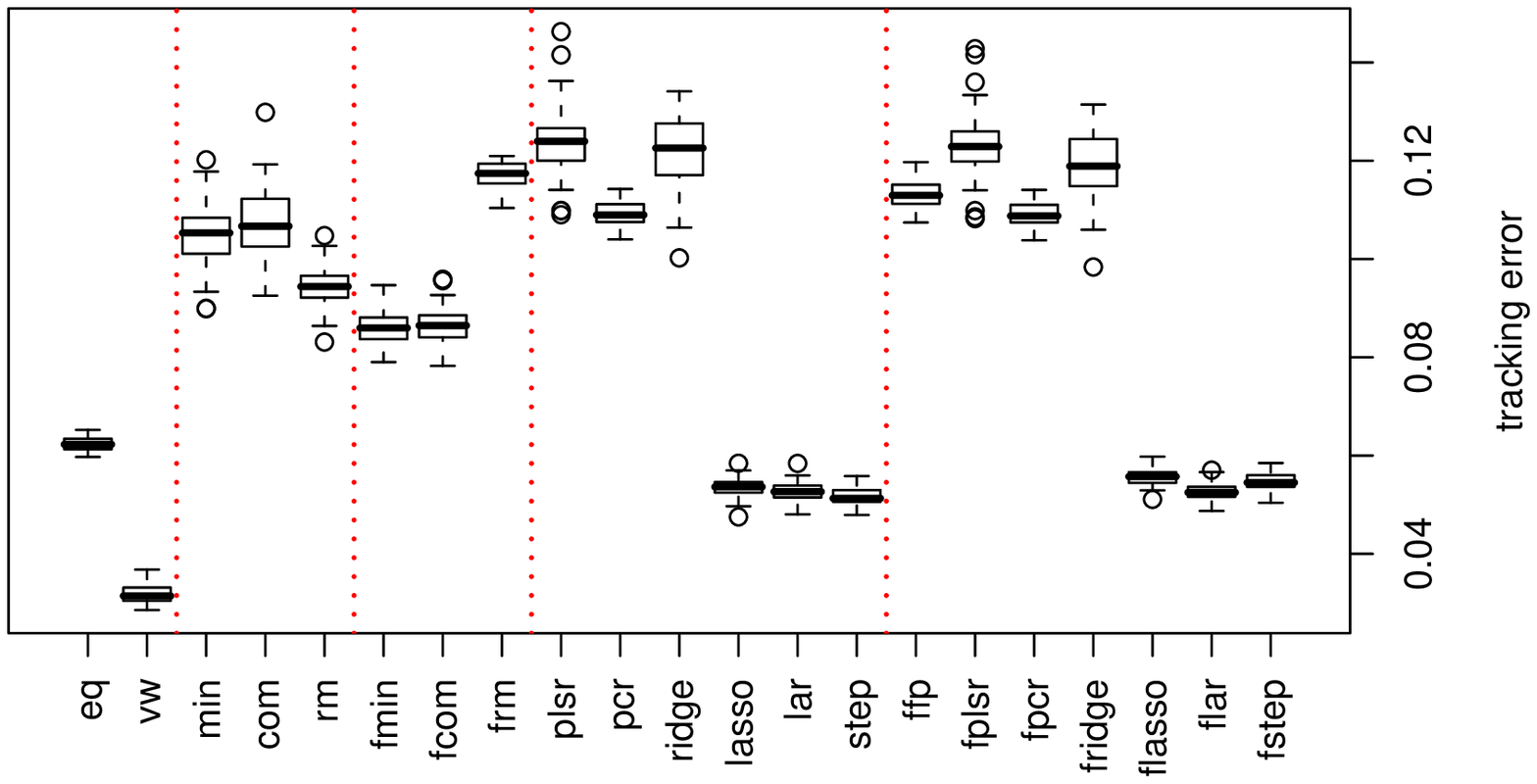}
\caption{Boxplots of Sharpe ratios {\em (top)} and the tracking error
  {\em (bottom)} obtained over 50 random paths through the 26 years,
  obtained by randomly sampling 250 qualifying assets in each year.
  The averages of these numbers is what is reported in Table
  \ref{t:sharpe}.  The horizontal bars correspond to the vertical ones
  in that table.}
\label{f:boot}
\end{figure}
Figure \ref{f:boot} compliments Table \ref{t:sharpe} by showing the
distribution (via boxplots) of the Sharpe ratios and the tracking
error obtained for each of the 50 random paths through the 26 years.
Recall that these were obtained by randomly sampling 250 qualifying
assets in each year.  The numbers in Table \ref{t:sharpe} are the
means of data use to construct each boxplot, whereas the boxplots in
the figure represent Monte Carlo approximations to the sampling
distribution of portfolio characteristics under the various estimators
of $\hat{\bm{\Sigma}}$.  In short, the figure reinforces the
superiority of the LARS estimators which, in addition to having large
Sharpe ratios and small tracking error, also exhibit small variability
with respect to Monte Carlo resampling.  It is interesting to note
that the LARS based estimators (without the factor) show the lowest
variability in their Sharpe ratios amongst all {\tt monomvn}
estimators.

It may be tempting to conclude that these results contradict the
results of the ELL--based comparison(s) on synthetic data in Section
\ref{sec:synth}.  Indeed, in that section we saw that PCR seemed to be
the best at recovering the (known) of the distribution which generated
the training data.  However, means, variances, Sharpe ratios, tracking
error, etc., are specific statistics, and moreover they are obtained
after a (highly non--linear) transformation into portfolio weights via
quadratic programming.  Therefore, we should expect to see different
results, since these statistics represent utilities which are
different from ELL.  That being said, notice that PCR is still the
best in terms of average annualized standard deviation (and thus
Sharpe ratio) [see Table \ref{t:sharpe}] when no appropriate factors
are available---but with high variability [see Figure \ref{f:boot}].
Importantly, both experiments (here and in Section \ref{sec:synth})
show, resoundingly, that using all of the available data via {\tt
  monomvn} is preferred over a complete data estimator.

\subsection{Examining dependence relationships between assets}
\label{sec:depend}

For our final empirical analysis we shall demonstrate the descriptive
power of {\tt monomvn}.  At the same time we shall take the
opportunity to show how the method can be applied when there are
thousands of assets.

From Thomson Financial's Datastream ({\tt www.datastream.com}), we
have downloaded, in dollar terms, the total returns data of each stock
in the Russell 3000$^{\mbox{\tiny \textregistered}}$
Index 
  representing the broad United States equity universe encompassing
  approximately 98\% of the market:
  1792 weekly returns between 12/01/1973 and 11/05/2007 for 2894
  assets. In order to obtain a set of clean and complete data, each
  series is tested for illiquidity, completeness, and stationarity,
  using the following methodology.  We removed assets which were
  marked to market at a frequency other than weekly, to exclude
  illiquid assets that may exhibit artificial serial correlation (this
  essentially excludes any stock that has more than two weeks of
  consecutive unchanging prices at any point in time).  Then, an
  augmented Dickey Fuller test \citep{dickey:fuller:1979} is employed
  to exclude any of the assets that exhibit non--stationarity (six
  lags have been tested at the 99\% confidence level).  A total of
  2461 stocks remained after applying these two filtering steps.
  There are 558 assets with longest history of 1792 returns; the least
  observed asset has only 76 returns (so the ``complete'' estimator(s)
  can use only 3\% of the data); the overall proportion of missing
  observations was 0.472.

We consider applying the lasso version of the {\tt monomvn} algorithm
to this data, with $p = 0$, i.e., always use the lasso (never use
OLS).  As we have mentioned, the lasso (and other LARS methods) have
descriptive (as well as predictive) power because they can provide
$\hat{\bm{\beta}}$ with many coefficients set to zero.  In the context
of the {\tt monomvn} algorithm this means that the MLE
$\hat{\bm{\Sigma}}$ may have zero entries, indicating marginally
uncorrelated assets, and moreover may have block--diagonal structure
(or zeros in $\hat{\bm{\Sigma}}^{-1})$ indicating a pairwise
conditional independence of assets.  Since ridge regression, PCR, and
PLSR always yield $|\hat{\beta}_i| > 0$, they would never produce a
zero in $\hat{\bm{\Sigma}}$ or $\hat{\bm{\Sigma}}^{-1}$, and so would
be less useful for creating such qualitative summaries of the
relationships between asset returns.  It may be tempting to interject
zeros where there are small values in $\hat{\bm{\Sigma}}$ or
$\hat{\bm{\Sigma}}^{-1}$, but like the ``complete'' and ``observed''
estimators, the resulting matrix would not usually be positive
definite.  Moreover, classical pairwise tests for independence, say
via the Pearson product--moment correlation coefficient, would give
unrealistic results.  With return histories as short as $\sim80$ weeks
and estimated correlation less than about 0.2, a simple calculation
shows that there would not be enough evidence to reject the
hypothesis that the correlation is zero.

The estimator obtained using the lasso on this data yields a
$\hat{\bm{\Sigma}}$ with 36\% of its entries set to zero.  Moreover,
50 of its 2641 columns (or 2\%) are everywhere zero except in the
diagonal position.  This means that 36\% of asset pairings are
marginally uncorrelated.  Investigating pairwise correlation between
assets, conditional on all of the others, involves looking for zeros
in $\hat{\bm{\Sigma}}^{-1}$, of which we find 140 (or 6\%).  This
means that the rows/columns of $\hat{\bm{\Sigma}}$ can be reordered so
that the matrix has block--diagonal structure, and that the returns of
6\% of the assets are conditionally independent.
\begin{figure}
\includegraphics[angle=-90,scale=0.8]{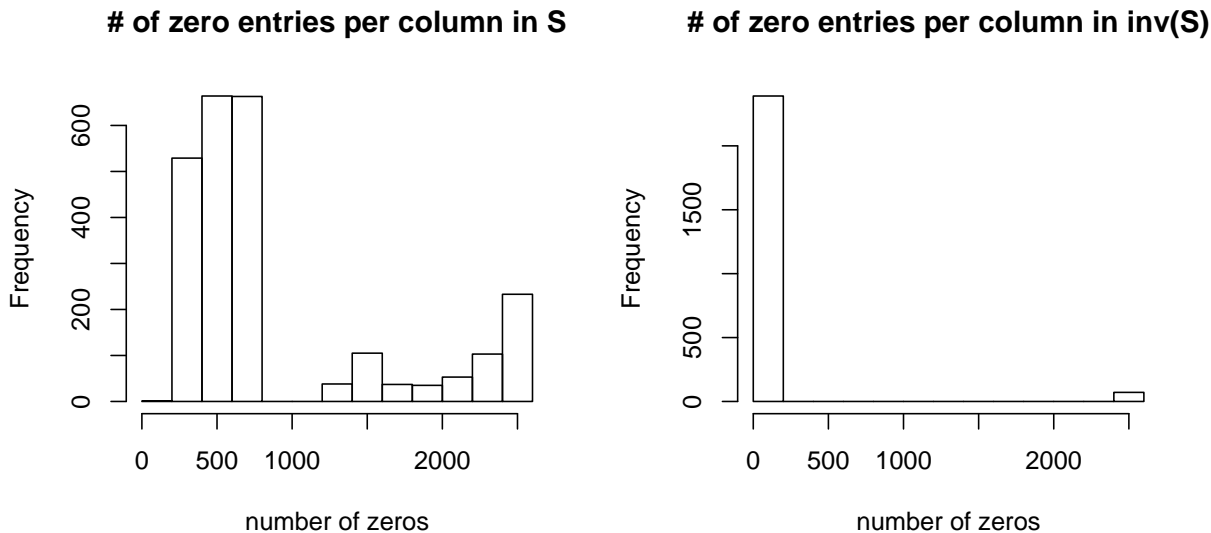}
\vspace{-0.1cm}
\caption{Histograms of the number of zeros in each column of
  $\hat{\bm{\Sigma}}$ {\em (left)} and $\hat{\bm{\Sigma}}^{-1}$ {\em
    (right)}.}
\label{f:indep}
\end{figure}
Figure \ref{f:indep} shows histograms summarizing the number of zeros
in each column of $\hat{\bm{\Sigma}}$ and $\hat{\bm{\Sigma}}^{-1}$.
Every column in both matrices had at least one zero entry.  The figure
clearly illustrates that the resulting correlations can be used to
cluster the assets, but this is beyond the scope of this paper.

To wrap up the experiment we downloaded the market returns available
from the Russel 3000 index 
for 1479 (of 1792) contiguous weeks ending 11/5/2007 and used them to
create a residual return series
for each of the 2461 assets in our
study.  We then re-ran the lasso experiment, above, to discover that
58\% of the asset parings are marginally uncorrelated and 14\% are
conditionally independent when the market is taken into account.  The
histograms corresponding to this experiment are similar to those for
the initial one, in Figure \ref{f:indep}, and so they are not
reproduced here.  

\section{Discussion}
\label{sec:discuss}

We have shown how the methods of \cite{stambaugh:1997} can be applied
for large numbers of assets whose histories are (nearly) unconstrained
in length.  The key insight is in replacing OLS regressions with more
parsimonious ones that either use derived input directions or apply
some sort of shrinkage.  Whereas Stambaugh demonstrated his
methodology on 22 assets, we have shown how the {\tt monomvn}
algorithm---essentially the same methodology with a different
regression method---can handle thousands.  We argued that even when
OLS regressions suffice, the more parsimonious ones can offer
improvements in both accuracy and interpretation.  We also argued that
it is advantageous to let a model selection method (e.g., parsimonious
regression) decide which dependencies between factors and returns
exist, as opposed to assuming a classical factor model structure.

\cite{stambaugh:1997} showed that by applying the standard
noninformative prior $\pi(\bm{\theta}) \propto
|\bm{\Sigma}|^{\frac{p-1}{2}}$ \citep[e.g.][pp.~154]{schafer:1997} it
is possible to turn the MLEs $\hat{\bm{\mu}}$ and $\hat{\bm{\Sigma}}$
into moments $\tilde{\bm{\mu}}=\hat{\bm{\mu}}$ and
$\tilde{\bm{\Sigma}}\ne\hat{\bm{\Sigma}}$ of a Bayesian posterior
(predictive) distribution that, when used in the mean--variance
framework, are said to take {\em estimation risk} into account.  We
note that, due to the notation used in that paper, it is a common
misconception that these posterior moments forecast the ML estimates
into the future.  Since Stambaugh employs the i.i.d.  assumption in
the same way that we do in Eq.(\ref{eq:iidlik}), these are only
moments of the posterior for $\bm{\theta}$ conditioned on the
available historical data.  Therefore, time is irrelevant, so the
moments apply to the past as well without modification.  Finally, to
label this approach as ``Bayesian'' is an overstatement.  While
Stambaugh is correct to note that estimates of the mean vector and
covariance matrix are all that are needed within the mean--variance
framework, what results is a point--estimate (vector) of optimal
portfolio weights, not (samples from) a Bayesian posterior
distribution, as would be ideal.  The challenge is that while the
moments of the posterior have a nice closed form, the distribution
itself does not.  Further challenges limit the application of this
approach in the ``big $p$ small $n$ setting''.  In this situation the
standard noninformative prior leads to an improper posterior.  This
can be most easily seen in the calculation of Stambaugh's $\tilde{V}
\equiv \tilde{\bm{\Sigma}}$ (in our notation) in Eq.~(69--71),
pp.~302, where the resulting diagonal would be negative.

Stambaugh's Bayesian approach is not the only way forward.  It is
possible to obtain the sampling covariance matrix of $\hat{\bm{\mu}}$
analytically.  However, an analytic form for the sampling variability
of $\hat{\bm{\Sigma}}$ is not known.  The bootstrap
\citep[e.g.][Sections 7.11 \& 8.2]{hastie:tibsh:fried:2001} offers a
Monte Carlo method for quantifying the {\em stability} of
$\hat{\bm{\Sigma}}$ via its component-wise confidence intervals.  We
took a related approach at the end of Section \ref{sec:portfolio} to
examine how variability in $\hat{\bm{\Sigma}}$, arising from random
subsamples of 250 assets, filters through to the properties of the
balanced portfolios.  However, \citet[][Section
7.4.4]{little:rubin:2002} make a strong argument in preference for a
fully Bayesian approach instead.  Facilitating tractable Bayesian
estimation for parsimonious regression algorithms, as would be
required by {\tt monomvn}, presents a serious challenge.  The Bayesian
lasso \citep{park:casella:2008} and so--called Bayesian latent factor
models \citep{west:2003}, which can be seen as a Bayesian extension of
principal components and partial least squares regressions, have
received much attention in the recent literature.  Exploring the
extent to which these can be applied within the {\tt monomvn}
algorithm to get samples from the posterior distribution of $\bm{\mu}$
and $\bm{\Sigma}$ is part of our ongoing work.  These samples can
accurately reflect the estimation risk in mean--variance portfolio
allocation by filtering the uncertainty though the optimization to get
a distribution on the simplex of portfolio weights.

Another interesting extension would involve relaxing the assumption of
(multivariate) normality, i.e., to decouple the dependence
distribution, or {\em copula} \citep{sklar:1957}, from the marginals.
In this regard, \cite{patton:2006} has made promising inroads into
applying copulas to a pair of return series under a monotone
missingness pattern.  Although the theory for copulas
\citep{nelsen:1999} naturally extends beyond two dimensions, the
application of the methodology quickly becomes intractable without
enforcing severely restrictive assumptions.  Our ongoing work includes
identifying ways in which the {\tt monomvn} algorithm for
high--dimensional estimation under monotone missingness may be
extended to support marginal Student--$t$ distributions and GARCH
models with various parametric forms of the copula.  While there is
plenty of evidence in the literature against the assumption of
normality for asset returns \citep[e.g.][]{mills:1927}, we argued that
the most important thing is to be able to make use of all of the
available data with an algorithm that is computationally tractable.


\bibliography{corr}
\bibliographystyle{jasa}

\end{document}